\documentclass[aps,prc,twocolumn,groupedaddress,showpacs,10pt]{revtex4-1}

\usepackage[utf8]{input enc}
\usepackage{amsfonts}
\usepackage{amsmath}
\usepackage{amssymb}
\usepackage[pdftex]{graphicx}

\begin{document}

\newcommand{\tj}[6]{ \begin{pmatrix}
  #1 & #2 & #3 \\
  #4 & #5 & #6
 \end{pmatrix}}


\title{Analytical transformed harmonic oscillator basis for
  \\three-body nuclei of astrophysical interest: Application to
  $^6$He} 




\author{J. Casal}
\email{jcasal@us.es}
\author{M. Rodr\'{\i}guez-Gallardo}
\email{mrodri@us.es}
\author{J.M. Arias}
\email{ariasc@us.es}
\affiliation{Departamento de F\'{\i}sica At\'omica, Molecular y Nuclear,
  Facultad de F\'{\i}sica, Universidad de Sevilla, Apartado 1065, E-41080
  Sevilla, Spain}


\date{\today}

\begin{abstract}
Recently, a square-integrable discrete basis, obtained performing a
simple analytical local scale transformation to the harmonic oscillator
basis, has been proposed and successfully applied to study the properties of
two-body systems. Here, the method is generalized
to study three-body systems. To test the goodness of the formalism and 
establish its applicability and limitations, the capture reaction rate for the
nucleosynthesis of the Borromean nucleus $^6$He ($^4\text{He}+n+n$) is
addressed. Results are compared with previous publications and with
calculations based on actual three-body continuum wave functions, which can be
generated for this simple case. The obtained
results encourage the application to other Borromean nuclei of
astrophysical interest such as $^9$Be and $^{12}$C, for which actual  three-body continuum calculations are very involved.
\end{abstract}


\maketitle



  \section{Introduction}

The study of three-body Borromean nuclei is known to be important
for astrophysical questions such as stellar
nucleosynthesis. Borromean nuclei are three-body systems whose
binary subsystems  
are unbound~\cite{Zhukov93}. One of the Borromean nuclei which has
attracted more interest is $^{12}$C $(\alpha+\alpha+\alpha)$ due to
the relevance of the triple-$\alpha$ reaction in the red giant phase of  
stars~\cite{Hoyle}. This process allows the formation of heavier
elements in stars, where mainly $\alpha$ particles and nucleons are
present, overcoming the $A=5$ and $A=8$ instability  
gaps~\cite{Aprahamian05}. The production rate of such process has not
yet been determined accurately for the entire temperature range
relevant in astrophysics~\cite{Nguyen1}. This is due to  
experimental problems to measure these processes as well as to
discrepancies in the theoretical predictions about the structure of
$^{12}$C. The formation of $^{12}$C has traditionally been  
studied as a sequential
process~\cite{Efros96,Sumiyoshi02,Bartlett06}. But, at low temperatures,
the three $\alpha$ particles have no access to intermediate resonances
and therefore they fuse  
directly~\cite{Nguyen1}. The description of this process requires an
accurate three-body model. 

Other Borromean nuclei are also important for nucleosynthesis in different
astrophysical scenarios. For instance, massive stars usually end up
with the explosion of a supernova and the possible  
formation of a neutron star. These neutron-rich environments, with
low density and high temperature (\textit{hot bubbles}), 
are an ideal medium for nucleosynthesis by rapid neutron capture
(or \textit{r}-process)~\cite{Meyer92,Meyer94}. Among these processes one finds the formation
of $^6$He $(\alpha + n+n)$ or $^9$Be  
$(\alpha +\alpha +n)$ that could also overcome the \mbox{$A=5,8$}
gaps~\cite{Bartlett06}.  Therefore, as in the case of the
triple-$alpha$ capture, understanding of these processes requires a very
accurate description of the states of $^{6}$He and $^9$Be in a three-body model
as well as the corresponding electromagnetic transition probabilities.  

In particular, the $^6$He nucleus has a halo structure. The halo nuclei are weakly-bound exotic systems in which one or more particles have a large probability of  being at distances far away from typical nuclear radii~\cite{Hansen95}. 
A common characteristics of these systems is
their small separation energy and hence their large breakup
probability. This process can be understood as an excitation of the nucleus to
unbound or scattering states that form a continuum of
energies~\cite{JALay10}. For that reason, the study of weakly bound
three-body systems, such as $^6$He, demands a proper treatment of the three-body problem with a reasonable description
of their continuum structure. In this work, a method, which includes these characteristics, is proposed and then applied to $^6$He as a benchmark calculation.
 It is worth noting that more fundamental few-body methods can be applied to  $^6$He considered as a six-nucleon system, such as the Resonating Group~\cite{Tang78} or the Lorentz Integral Transform~\cite{Efros94,Bacca02} methods.


From the theoretical point of view, the treatment of unbound  
states of a quantum-mechanical system deals with the drawback that the corresponding wave functions are not square-normalizable and their energies are not discrete values. Solving this problem is a difficult task, especially as the number of charged particles increases, since one needs to know the asymptotic behavior of the unbound states. 
Nevertheless, there are various procedures to address this problem such as the R-matrix method~\cite{Wigner47,IJThompson00,Desc06}, not without difficulties. 
Another approach to solve the continuum problem consists in using the so-called discretization methods. These methods replace the true continuum by a finite set of normalizable states,
i.e., a discrete basis that can be truncated to a relatively 
small number of states and nevertheless provide a reasonable
description of the system. Several   
discretization methods have been proposed~\cite{Zhukov93}.  
For instance, one can solve the Schr\"odinger equation in a box~\cite{RdDiego10}, being the energy level density  governed by the size of the box. As this is larger, the energy level density increases but numerical problems begin to appear. Another method is the binning procedure, used traditionally in the
Continuum-Discretized Coupled-Channels (CDCC) formalism~\cite{Austern87}. In
this method the continuum spectrum is truncated at a maximum energy
and divided into a finite number of energy intervals or bins. For each bin, a
normalizable state is constructed by superposition of the scattering states 
within that interval. This approach requires, first the calculation of the unbound states and then the matching with the correct asymptotic behavior.
As mentioned above, the calculation of this asymptotic
behavior for a three-body system with charged particles is by no means
an easy task. 

An alternative method to obtain a discrete representation of
the continuum spectrum is the pseudostate (PS) method, which consists in  
diagonalizing the three-body Hamiltonian in a complete set of
$\mathcal{L}^2$ wave functions (that is, square integrable). The eigenstates 
of the Hamiltonian are then taken as a discrete representation of  
the spectrum of the system. The advantage of this procedure is that it does
not require going through the continuum wave functions and the 
knowledge of the asymptotic behavior is not needed. 
A variety of bases have been proposed for two-body~\cite{HaziTaylor70,Matsumoto03,MRoGa04,AMoro09} and also
for three-body calculations~\cite{Matsumoto04,Rasoanaivo89,MRoGa05}. 

In previous works, a PS method based on a local scale transformation (LST) 
of the harmonic oscillator (HO) basis has been proposed~\cite{FPeBe01}. 
When the ground state of the system is known, a  
useful procedure to discretize the continuum consists in performing a
numerical LST that transforms the actual ground-state wave function of 
the system into the HO ground state. Once the LST is obtained,  
the inverse transformation is applied to the HO basis, giving rise to
the transformed harmonic oscillator (THO) basis. This method has been
used to describe the two-body continuum in  
structure~\cite{MRoGa04} and reactions~\cite{AMoro01,AMoro06} studies,
showing that the THO method together with the CDCC technique is useful
to describe continuum effects in nuclear  
collisions. The method was also applied to
$^6$He~\cite{MRoGa05,MRoGa08}, showing that the numerical THO method
is appropriate to describe three-body weakly bound systems with a
relative small THO basis.  
In most recent works~\cite{AMoro09,JALay10} an alternative
prescription to define the LST was proposed, introducing an analytical
transformation taken from Karataglidis  
\textit{et al.}~\cite{Karataglidis}. This analytical transformation
keeps the simplicity of the HO functions, but converts their Gaussian
asymptotic behavior into an exponential one, more  
appropriate to describe bound systems. This analytical THO method has been
applied to study two-body systems, providing a suitable representation
of the bound and unbound spectrum to calculate  
structure and scattering observables within the CDCC method~\cite{AMoro09}. 
The analytical THO presents several advantages over the numerical THO. 
(1) It is not needed to know previously the ground-state wave function of
the system considered. (2) Due to the analytical form of the
transformation, it can be easily implemented in a numerical code. 
(3) The parameters of the transformation govern the radial extension 
of the THO basis allowing the construction of an optimal basis for 
each observable of interest.

In this work, we extend the analytical THO method to study
three-body systems. We start with the
construction of the basis, then we diagonalize the three-body
Hamiltonian, and we compute the transition probabilities needed
for the calculation of the reaction rate. As a simple
example of application, we check the formalism  for the
Borromean nucleus $^6$He. For this, a rich variety of  
data is available~\cite{Aksough03,Egelhof03,Aumann98,Aumann99,Aguilera01,Aguilera00,Kakuee03}
and can be used to benchmark theoretical models. Finally, the
structure calculation allows us to determine the rate  
of the radiative capture reaction \mbox{$^4\text{He}+ n + n
  \rightarrow$ $^6\text{He} + \gamma$}.
It is known  this reaction is not of great astrophysical interest but provides 
a robust test for our three-body model. In this case, with just one charged particle, one can generate easily the continuum wave functions and our model
 results can be confronted to actual continuum calculations.  The study of this reaction will validate our formalism so as to make it reliable when applied to cases in 
which such comparisons with the true continuum cannot be easily done. This will be the case of $^9$Be, $^{12}$C, or $^{17}$Ne, that are subjects for future research.

The manuscript is structured as follows. In Sec. \ref{sec:THO} the
analytical THO method for three-body systems is completely worked out: 
basis, matrix elements, and calculation of transition probabilities. In
Sec. \ref{sec:rate} the expressions and concepts involved in the  
calculation of the radiative capture reactions of three particles 
into a bound nucleus are discussed. In
Sec. \ref{sec:application} the full formalism is applied to the case
of $^6$He. Finally, in Sec. \ref{sec:conclusions}, the main
conclusions of this work are summarized.  


\section{PS Method: Analytical THO for three-body systems}\label{sec:THO}

\begin{figure}
\includegraphics[width=0.5\linewidth]{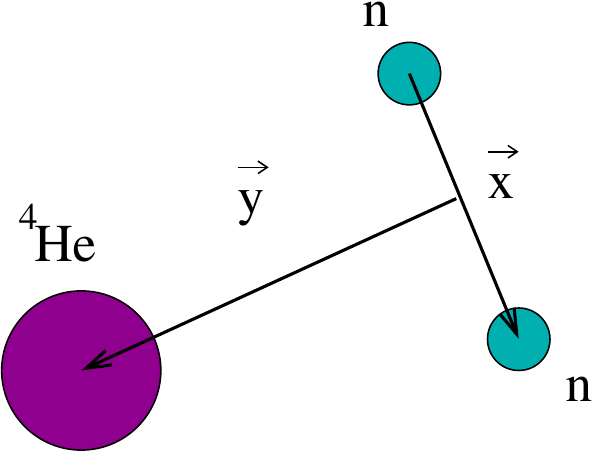}
\caption{(Color online) The Jacobi $T$-coordinate system used to describe the $^6$He nucleus.}
\label{fig:Jacobi}
\end{figure}

Jacobi coordinates $\{\boldsymbol{x},\boldsymbol{y}\}$, illustrated in
Fig.~\ref{fig:Jacobi}, are used to describe three-body systems [six-dimensional problems]. The variable $\boldsymbol{x}$ is proportional to the
relative coordinate between two of the particles and $\boldsymbol{y}$
is proportional to  
the coordinate from the center of mass of these two particles to the
third one, both with a scaling factor depending on their
masses~\cite{MRoGa05}. Please, note that there are three different Jacobi
systems. From the Jacobi coordinates, one can  
define the hyperspherical coordinates
$\{\rho,\alpha,\widehat{x},\widehat{y}\}$, where $\rho=\sqrt{x^2+y^2}$
is the hyper-radius and $\tan\alpha=x/y$ defines the hyperangle.  

The PS method consists in diagonalizing the Hamiltonian of the system of
interest in a discrete
basis of $\mathcal{L}^2$ functions. Using hyperspherical coordinates,
and introducing  
$\Omega\equiv\{\alpha,\widehat{x},\widehat{y}\}$ for the angular
dependence, the state wave functions of that basis can be expressed as 
\begin{equation}
 \psi_{i\beta j\mu} (\rho,\Omega) = R_{i\beta}(\rho) \mathcal{Y}_{\beta j\mu}(\Omega).
\label{eq:basis}
\end{equation}
Here $\mathcal{Y}_{\beta j\mu}(\Omega)$ are states of good total
angular momentum, expanded in hyperspherical harmonics
(HH)~\cite{Zhukov93,MRoGaTh} $\Upsilon_{Klm_l}^{l_xl_y}(\Omega)$ as 
\begin{eqnarray}
\mathcal{Y}_{\beta j\mu}(\Omega)& =& \sum_{\nu\iota}\langle j_{ab}\nu I\iota|j\mu\rangle \kappa_I^{\iota} \nonumber \\
&\times&  \sum_{m_l\sigma}\langle lm_lS_x\sigma|j_{ab}\nu\rangle
\Upsilon_{Klm_l}^{l_xl_y}(\Omega)\chi_{S_x}^{\sigma},
\label{eq:HHexpand}
\end{eqnarray}
and $\beta\equiv\{K,l_x,l_y,l,S_x,j_{ab}\}$  is a set of quantum
numbers called channel. In this set, $K$ is the hypermomentum, $l_x$ and  
$l_y$ are the orbital angular momenta associated with the Jacobi
coordinates $\boldsymbol{x}$ and $\boldsymbol{y}$, respectively, $l$  is the
total orbital angular momentum
($\boldsymbol{l}=\boldsymbol{l_x}+\boldsymbol{l_y}$), $S_x$ is the
spin of the particles related by the coordinate $\boldsymbol{x}$, and
$j_{ab}$ results from the coupling
$\boldsymbol{j_{ab}}=\boldsymbol{l}+\boldsymbol{S_x}$. If we denote by
$I$  the spin of the third particle, which we assume to be fixed, the
total angular momentum $j$ is $\boldsymbol{j}=\boldsymbol{j_{ab}} +
\boldsymbol{I}$. With that notation, $\chi_{S_x}^{\sigma}$ is  
the spin wave function of the two particles related by the Jacobi
coordinate $\boldsymbol{x}$, and $\kappa_I^{\iota}$ is the spin
function of the third particle. The HH are eigenfunctions of the
hypermomentum operator $\widehat{K}^2$, and can be expressed in terms
of the spherical harmonics as 
\begin{eqnarray}
\Upsilon_{Klm_l}^{l_xl_y}(\Omega)&=&\sum_{m_{x}m_{y}}\langle l_xm_xl_ym_{y}|lm_l \rangle \Upsilon_{K}^{l_xl_ym_{x}m_{y}}(\Omega),\\
 \Upsilon_K^{l_x l_y m_x m_y}(\Omega)&=&\varphi_K^{l_x l_y}(\alpha)
 Y_{l_x m_x}(\widehat{x}) Y_{l_y m_y}(\widehat{y}),\\ 
 \label{eq:HH}
\varphi_K^{l_x l_y}(\alpha) &=& N_K^{l_x l_y} (\sin\alpha)^{l_x}
(\cos\alpha)^{l_y} \nonumber \\
&\times& P_n^{l_x+\frac{1}{2},l_y+\frac{1}{2}}(\cos 2\alpha)
\end{eqnarray}
where $P_n^{a,b}$ is a Jacobi polynomial with order $n=(K-l_x-l_y)/2$ and $N_K^{l_x l_y}$ is the normalization constant.



On the other hand, $R_{i\beta}(\rho)$ are the hyperradial wave functions, where
the label $i$ denotes the hyperradial excitation. The form of these functions
depends on the PS method used. Then, the states of the system are given by
diagonalization of the three-body Hamiltonian in a finite basis up to $i_{\rm max}$
hyperradial excitations in each channel,
\begin{align}
\nonumber \Psi_{nj\mu}(\rho,\Omega)= & \sum_{\beta} \sum_{i=0}^{i_{max}} C_{n}^{i\beta j}\psi_{i\beta j\mu} (\rho,\Omega) \\
				   = & \sum_{\beta} \underbrace{\left(\sum_{i=0}^{i_{\rm max}} C_{n}^{i\beta j} R_{i\beta}(\rho)\right)}_{\mathcal{R}^{nj}_{\beta}(\rho)} \mathcal{Y}_{\beta j\mu}(\Omega),
\label{eq:eigenwf}
\end{align}
being $C_{n}^{i\beta j}$ the diagonalization coefficients and 
$\mathcal{R}^{nj}_{\beta}(\rho)$ the hyperradial wave function corresponding to the
channel $\beta$. The label $n$ enumerates the eigenstates. 

    \subsection{Analytical THO method}

As stated in the introduction, several PS bases have been proposed for 
three-body studies~\cite{Matsumoto04,Rasoanaivo89,MRoGa05,FPeBe01}. Here, we use the
THO method based on a LST of the HO 
functions, so the hyperradial wave functions are obtained as 
\begin{equation}
  R_{i\beta}^{\text{THO}}(\rho)=\sqrt{\frac{ds}{d\rho}}R_{iK}^{\text{HO}}[s(\rho)].
\label{eq:R}
\end{equation}
Note that, meanwhile the THO hyperradial wave functions depend, in general, on all the quantum numbers included in a channel $\beta$, the HO hyperradial  wave functions only depend on one of them, the hypermomentum $K$.
The transformation $s(\rho)$ is not unique, and in this work we adopt the analytical form
 of Karataglidis \textit{et al.}~\cite{Karataglidis},
\begin{equation}
s(\rho) = \frac{1}{\sqrt{2}b}\left[\frac{1}{\left(\frac{1}{\rho}\right)^{\xi} +
\left(\frac{1}{\gamma\sqrt{\rho}}\right)^\xi}\right]^{\frac{1}{\xi}},
\label{eq:LST}
\end{equation}
depending on the parameters $\xi$, $\gamma$, and the oscillator length $b$.
The HO hyperradial variable $s$ is dimensionless according to the
transformation defined above [Eq.~(\ref{eq:LST})]. In this way, we take the
oscillator length $b$ as another parameter of the transformation.

The function $s(\rho)$ behaves asymptotically as
$\frac{\gamma}{b}\sqrt{\frac{\rho}{2}}$  and hence the THO hyperradial 
wave functions 
obtained behave at large distances as $\exp{(-\gamma^2\rho/2b^2)}$. Therefore, 
the ratio $\gamma/b$ governs the asymptotic behavior of the THO functions: as
$\gamma/b$ increases, the hyperradial extension of the basis decreases and
some of the eigenvalues obtained by diagonalizing the Hamiltonian explore higher
energies~\cite{JALay10}. That is, $\gamma/b$ 
determines the density of PSs as a function of the energy. Concerning the 
parameter $\xi$, the authors of Ref.~\cite{Karataglidis} found a very weak 
dependence of the results on this parameter. Because of that, we have fixed for
all calculations $\xi=4$ as in Refs.~\cite{JALay10,AMoro09}.

\begin{figure}
\includegraphics[width=0.95\linewidth]{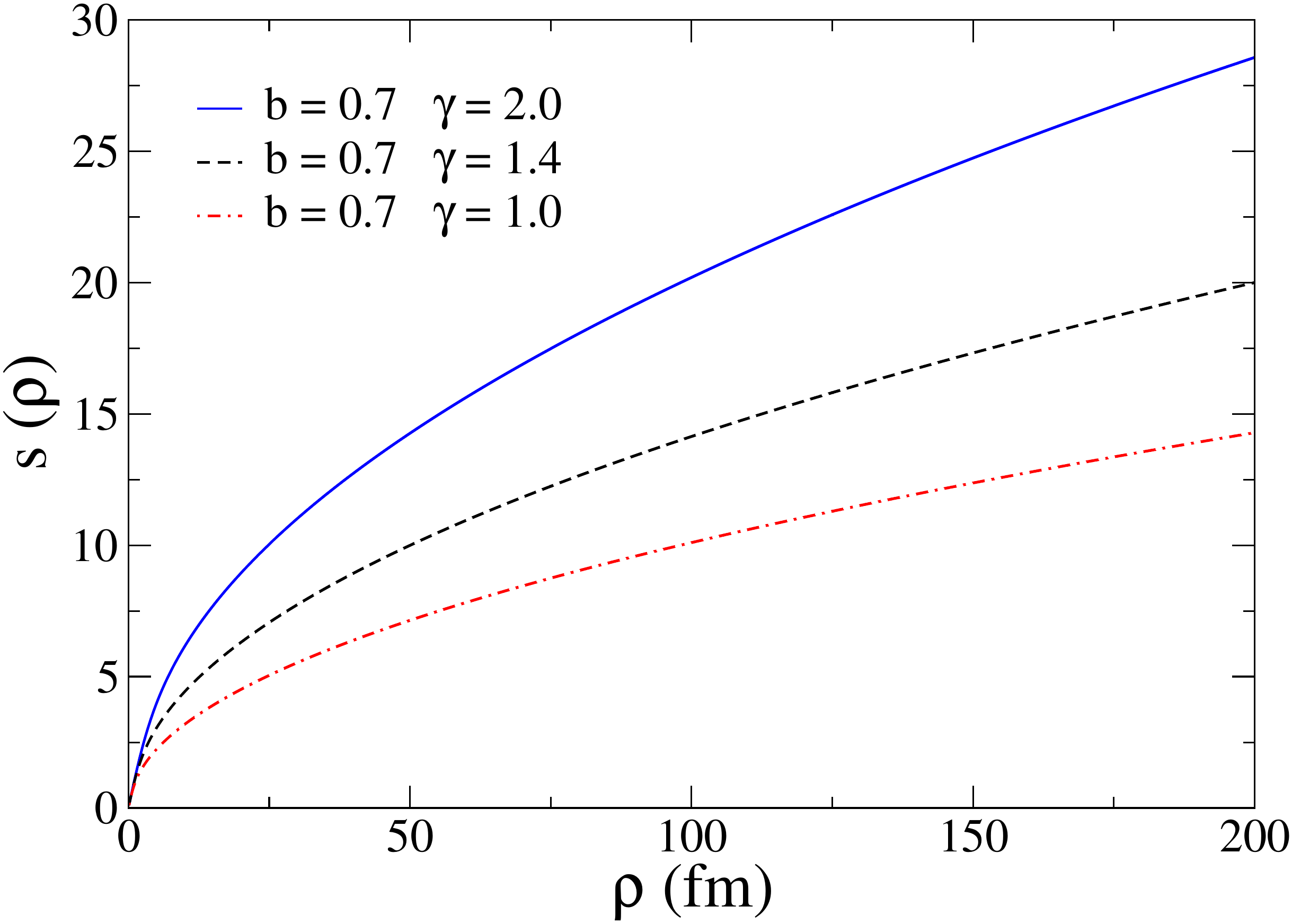}
\caption{(Color online) Different LSTs with parameter $b=0$.$7$ fm and three
values  of $\gamma$: 2.0, 1.4, and 1.0 fm$^{1/2}$.}
\label{fig:LST}
\end{figure}

The freedom to control the hyperradial extension of the THO basis is
an advantage of the analytical THO method. Depending on the observable of
interest, one is able to choose either a basis with a finer description of the 
low energy region (close to the breakup threshold) or a basis carrying more
information on the high energy spectrum. 
In Fig.~\ref{fig:LST} the LSTs for a fixed $b$ and different $\gamma$ values 
are presented.

    \subsection{Hamiltonian matrix elements}

The three-body Hamiltonian  in hyperspherical coordinates is written as
\begin{equation}
\widehat{\mathcal{H}}(\rho,\Omega) = \widehat{T}(\rho,\Omega) + \widehat{V}(\rho,\Omega).
\label{eq:H}
\end{equation}
The kinetic energy operator is~\cite{Nielsen01,MRoGa05}
\begin{equation}
\widehat{T}(\rho,\Omega) = -\frac{\hbar^2}{2m}\left[\frac{\partial^2}{\partial\rho^2}+\frac{5}{\rho}\frac{\partial}{\partial\rho}-\frac{1}{\rho^2}\widehat{K}^2(\Omega)\right],
\label{eq:T}
\end{equation}
where $m$ is a normalization mass that we take as the nucleon mass and 
$\widehat{K}^2(\Omega)$ represents the hyperangular momentum or hypermomentum
operator. $\widehat{T}(\Omega)$ does not 
connect different channels $\beta$ or states with different total angular
momentum $j$.  The Hamiltonian matrix elements have to be calculated between
states given by Eq.~(\ref{eq:basis}), which separates the hyperradial and
hyperangular parts. The hyperradial wave functions are constructed with
Eq.~(\ref{eq:R}) and satisfy the same normalization condition as the HO 
functions in six dimensions~\cite{MRoGaTh},
\begin{equation}
\int_0^\infty d\rho~\rho^5 R_{i\beta}^{\text{THO}}(\rho) R_{i'\beta}^{\text{THO}}(\rho) = \delta_{ii'}.
\end{equation}
For convenience, we introduce the hyperradial wave functions 
$U_{i\beta}^{\text{THO}}(\rho)$ as 
\begin{equation}
U_{i\beta}^{\text{THO}}(\rho)=\rho^{5/2}R_{i\beta}^{\text{THO}}(\rho),
\label{eq:U}
\end{equation}
which satisfy the orthonormality relationship
\begin{equation}
\int_0^\infty d\rho~U_{i\beta}^{\text{THO}}(\rho) U_{i'\beta}^{\text{THO}}(\rho) = \delta_{ii'}.
\end{equation}
With these functions, the kinetic energy operator can be re-written as
\begin{equation}
\widehat{T}_U(\rho)=-\frac{\hbar^2}{2m}\left[\frac{d^2}{d\rho^2}-\frac{15/4+K(K+4)}{\rho^2}\right]
\label{eq:Tu}
\end{equation}
and its matrix elements are~\cite{JCasalMSc}
  \begin{align}
  \nonumber & \langle i \beta j | \widehat{T}(\rho,\Omega) | i'\beta' j\rangle  = \langle  i \beta j | \widehat{T}_U(\rho) | i'\beta j\rangle ~ \delta_{\beta\beta'} \\
  \nonumber & = \delta_{\beta\beta'}~\frac{\hbar^2}{2m} ~\left[ \int_0^\infty d\rho ~\frac{dU_{i\beta}^{\text{THO}}(\rho)}{d\rho}~\frac{dU_{i'\beta}^{\text{THO}}(\rho)}{d\rho} \right. \\
  \nonumber& +\left(\frac{15}{4}+K(K+4)\right) \left. \int_0^\infty d\rho ~U_{i\beta}^{\text{THO}}(\rho)\frac{1}{\rho^2}U_{i'\beta}^{\text{THO}}(\rho) \right], \\
  & \left.\right.
  \end{align}
where the anti-hermiticity of the derivation operator has been taken into
account.  

The potential energy operator does connect, in general, different channels 
within the same $j$. The hyperangular integration is performed by using a set of
subroutines from the code 
FaCE~\cite{IJThompson04} that provides the hyperangular matrix elements 
$V_{\beta\beta'}^j(\rho)$, depending on $\rho$. These functions are then
integrated in the hyperradial variable, obtaining the potential energy matrix
elements as
\begin{eqnarray}
 \langle  i \beta j| \widehat{V}(\rho,\Omega) | i'\beta' j \rangle =\nonumber\\ \int_0^\infty d\rho ~ U_{i\beta}^{\text{THO}}(\rho) V_{\beta\beta'}^j(\rho) ~ U_{i'\beta'}^{\text{THO}}(\rho).
\end{eqnarray}

Once the kinetic energy and potential matrix elements are computed, the 
Hamiltonian is diagonalized in a truncated THO basis with $i_{\rm max}$ and the eigenstates of the system are obtained.

    \subsection{Transition probabilities $\boldsymbol{B(E\lambda)}$}

As in Ref.~\cite{MRoGa05}, we follow the notation of Brink and 
Satchler~\cite{BrinkSatchler}. The reduced transition probability between 
states of a system is defined as 
\begin{eqnarray}
 \nonumber B(E\lambda)_{nj,n'j'} & \equiv & B(E\lambda;nj\rightarrow n'j') \\
 & =& |\langle nj\|\widehat{Q}_\lambda\|n'j'\rangle|^2\left(\frac{2\lambda+1}{4\pi}\right),
\label{eq:BE}
\end{eqnarray}
where $\widehat{Q}_\lambda$ is the electric multipole operator of order 
$\lambda$. 

When a three-body system with only one charged particle, such  as
$^6$He ($^4\text{He}+n+n$), is considered and the Jacobi system {\bf T} illustrated 
in Fig.~\ref{fig:Jacobi} is used, the operator $\widehat{Q}_\lambda$ reads as
\begin{equation}
Q_{\lambda M_{\lambda}}(\boldsymbol{y})=\left(\frac{4\pi}{2\lambda+1}
\right)^{1/2}Z~e~\left(\frac{\sqrt{ma_y}}{m_c}\right)^\lambda y^{\lambda}
Y_{\lambda M_{\lambda}}(\widehat{y}).
\end{equation}
In this expression $Z$ is the atomic number of the system, $e$ is the electron 
charge, $m$ the mass of the nucleon, $a_y$ the reduced mass of the
subsystem related by the Jacobi coordinate $\boldsymbol{y}$ and $m_c$ the mass
of the charged particle (the core in the case of $^6$He).
The reduced matrix elements of this operator  can be expanded in terms of the 
THO basis obtaining the expression~\cite{MRoGaTh,MRoGa08}
\begin{eqnarray}
  \langle nj  \|\widehat{Q}_\lambda\|n'j'\rangle &=& (-1)^{j+2j'}\text{ } \hat{j}' Z\text{ } e \left(\frac{\sqrt{ma_y}}{m_c}\right)^\lambda\label{eq:Qlambda} \\
 & \times&  \sum_{\beta\beta'}  \delta_{l_x l_x'}\delta_{S_x S_x'}\delta_{j j_{ab}}\delta_{j' j_{ab}'} (-1)^{l_x + S_x}\nonumber \\
&\times& \hat{l}_y \hat{l}_y' \hat{l} \hat{l}' W(ll'l_yl_y';\lambda l_x)W(jj'll';\lambda S_x) \nonumber\\
&\times& \tj{l_y}{\lambda}{l_y'}{0}{0}{0} \sum_{ii'} C_n^{i\beta j} C_{n'}^{i'\beta' j'}\nonumber\\
&\times& \int\int d\alpha~d\rho~(\sin\alpha)^2(\cos\alpha)^2  \nonumber \\
& \times &  U^{\text{THO}}_{i\beta}(\rho)\varphi^{l_xl_y}_K(\alpha)y^\lambda\varphi^{l_x l_y'}_{K'}(\alpha)U^{\text{THO}}_{i'\beta'}(\rho).\nonumber
\end{eqnarray}
 
Since $n$ and $n'$ enumerate the different eigenstates, transition
probabilities  given by Eq.~(\ref{eq:BE}) are a set of discrete values. In order
to obtain continuous energy distributions 
from discrete values, the best option is to do the overlap 
with the continuum wave functions~\cite{MRoGa11}, if they are known. In this case the smoothed THO $B(E1)$ distribution must coincide perfectly with the actual continuum $B(E1)$ distribution. 
When the continuum states are not available, 
 it is  considered that, in general, a PS with energy
$\varepsilon_n$  is the superposition of continuum states in the vicinity. There
are several ways to assign an energy distribution 
to a PS~\cite{PDes12,Macias87}. In this work, for each discrete value of 
$B(E\lambda)(\varepsilon_n)$, a Poisson distribution
$D(\varepsilon,\varepsilon_n,w)$ with the following form is assigned,
\begin{equation}
  D(\varepsilon,\varepsilon_n,w)=\frac{(w+1)^{(w+1)}}{\varepsilon_n^{w+1}\Gamma(w+1)} \varepsilon^w\exp{\left(-\frac{w+1}{\varepsilon_n}\varepsilon\right)},
\label{eq:poisson}
\end{equation}
which is properly normalized. The parameter $w$ controls the width of the 
distributions; as $w$ decreases, the width of the distributions increases.
Finally, the $B(E\lambda)$ distribution is given by the expression
\begin{equation}
 \frac{dB(E\lambda)}{d\varepsilon}(\varepsilon,w)=\sum_n D(\varepsilon,\varepsilon_n,w)~ B(E\lambda)(\varepsilon_n).
\label{eq:distBE}
\end{equation} 
The $B(E1)$ distribution so obtained can be compared easily in the case of $^6$He with the continuum distribution in order to check the smoothing procedure.

One can also calculate the sum rules for electric transitions from the ground 
state (g.s.) to the states $(n,j)$ in order to test the completeness of the basis used. Using the Eq.~(\ref{eq:BE})
\begin{equation}
\sum_n
B(E\lambda)_{\text{g.s.},nj}=\left(\frac{2\lambda+1}{4\pi}\right)\sum_n|\langle 
\text{g.s.}\|\widehat{Q}_\lambda\|nj\rangle|^2,
 \label{eq:sumBE}
\end{equation}
a closed expression is obtained 
\begin{equation}
\sum_n
B(E\lambda)_{\text{g.s.},nj}=\frac{2\lambda+1}{4\pi}\frac{Z^2e^2m^\lambda
  a_y^\lambda}{m_c^{2\lambda}}\langle \text{g.s.}| y^{2\lambda}
|\text{g.s.}\rangle. 
 \label{eq:sumrule}
\end{equation}

  \section{Radiative capture reaction rate}\label{sec:rate}

The formalism introduced above allows calculations of astrophysical interest.
As stated in the introduction, some Borromean nuclei are important in the
nucleosynthesis processes, and an accurate knowledge of their reaction and 
production rates in different scenarios is essential to understand the origin 
of the different elements in the Universe. We focus on radiative capture
reactions of three particles, ($abc$), 
into a bound nucleus $A$ of binding energy $|\varepsilon_B|$, i.e., 
\mbox{$a+b+c \rightarrow$ $A + \gamma$}. The energy-averaged reaction rate for
such process, 
$\langle R_{abc}(\varepsilon)\rangle$, is given as a function of the temperature
 $T$ by the expression~\cite{Garrido11}
\begin{equation}
 \langle R_{abc}(\varepsilon) \rangle(T) = \int
R_{abc}(\varepsilon)F_B(\varepsilon,T) d\varepsilon.
\label{eq:aREabc}
\end{equation}
The function $F_B(\varepsilon,T)$ is the Maxwell-Boltzmann distribution and 
$R_{abc}(\varepsilon)$ is the radiative capture reaction rate at a certain
excitation energy $\varepsilon$.
It can be obtained from the inverse photodissociation
process~\cite{RdDiego10,Garrido11}  and is given by the expression
\begin{equation}
 R_{abc}(\varepsilon)=\nu!\frac{\hbar^3}{c^2}\frac{8\pi}{(a_x a_y)^{3/2}}\left(\frac{\varepsilon_\gamma}{\varepsilon}\right)^2 \frac{2g_A}{g_a g_b g_c}\sigma_\gamma(\varepsilon_\gamma),
\label{eq:REabc}
\end{equation}
where $\varepsilon=\varepsilon_\gamma+\varepsilon_B$ is the initial three-body 
kinetic energy, $\varepsilon_\gamma$ is the energy of the photon emitted, $\varepsilon_B$ is the ground-state energy, $g_i$
are the spin degeneracy of the 
particles, $\nu$ is the number of identical particles in the three-body system, 
and $a_x$ and $a_y$ are the reduced masses of the subsystems related to the
Jacobi coordinates 
$\{\boldsymbol{x},\boldsymbol{y}\}$. The photodissociation cross section 
$\sigma_\gamma(\varepsilon_\gamma)$ of the nucleus $A$ can be expanded into
electric and magnetic multipoles~\cite{RdDiego10,Forseen03}
\begin{equation}
\sigma_\gamma^{(\mathcal{O}\lambda)}(\varepsilon_\gamma)=\frac{(2\pi)^3 (\lambda+1)}
{\lambda[(2\lambda+1)!!]^2}\left(\frac{\varepsilon_\gamma}{\hbar
c}\right)^{2\lambda-1}\frac{dB(\mathcal{O}\lambda)}{d\varepsilon},
\label{eq:xsection}
\end{equation}
 which are related to the transition probability distributions
$dB(\mathcal{O}\lambda)/d\varepsilon$, for $\mathcal{O}=E, M$. 

From Eqs.~(\ref{eq:aREabc}) and (\ref{eq:REabc}),
we write the energy-averaged capture reaction rate expression for the  
contribution of order $\lambda$ as
\begin{align}
 \nonumber\langle R_{abc}(\varepsilon) \rangle(T) = & ~\nu!\frac{\hbar^3}{c^2}\frac{8\pi}{(a_x a_y)^{3/2}}\frac{g_A}{g_a g_b g_c} \frac{1}{(k_B T)^3}  \\
 \times & \int_0^\infty (\varepsilon+|\varepsilon_B|)^2 \sigma_\gamma^{(\mathcal{O}\lambda)}(\varepsilon+|\varepsilon_B|) e^{\frac{-\varepsilon}{k_B T}} d\varepsilon.
 \label{eq:aRE}
\end{align}
This integral is very sensitive to the
$dB(\mathcal{O}\lambda)/d\varepsilon$ behavior at low energy and, for 
that reason, a detailed description of the transition probability distribution in that region is needed to avoid numerical errors.
 Accordingly to the traditional literature~\cite{Weiss}, in absence of low energy resonances, the first multipole contribution is the dominant one and the electric contribution dominates over the magnetic one at the same order.

  \section{Application to $^6$He}\label{sec:application}

The $^6$He nucleus can be explained as a three-body system, formed by
an inert $\alpha$ core and two valence neutrons. This is the simplest case to
test the formalism developed in this work since there is just one charged
particle and the three-body continuum wave functions can be generated easily.
Comparison with actual continuum wave functions may serve as a
reference for any other calculation. In addition, valuable experimental
information is available on the ground state:
total angular momentum $j^\pi=0^+$, experimental binding energy of 0.975 MeV~\cite{Brouder12}, and rms point nucleon matter radius within 2.5$-$2.6
fm~\cite{Danilin05}. It has also a well-known $2^+$ resonance at 0.824
MeV over the breakup threshold.

To describe $^6$He, we use a model Hamiltonian that includes the two-body 
$n$-$n$ and $\alpha$-$n$ potentials, and also a simple central
hyperradial three-body force. These potentials are those used in
Ref.~\cite{MRoGa05}; the $n$-$\alpha$ potential taken from
Refs.~\cite{Bang79,IJThompson00}, with central and spin-orbit
components, and the GPT $n$-$n$ potential~\cite{GPT} with central,
spin-orbit, and tensor components. These two-body potentials are kept fixed for any total angular momentum and parity $j^{\pi}$.
However, this Hamiltonian does not include
all possible potential contributions. To include them effectively, a three-body force is usually introduced. In this work we have used the simple power form 
\begin{equation}
V_{3b}(\rho)= \frac{v_{3b}}{1+\left(\frac{\rho}{r_{3b}}\right)^{a_{3b}}}.
\label{eq:3b_force}
\end{equation}
The parameters $v_{3b}$, $r_{3b}$, and $a_{3b}$ have been  chosen to adjust the energy of the $0^+$ ground state and the position of the known $2^+$ resonance to the experimental values.

In three-body models of halo nuclei, such as $^6$He, the Pauli
principle treatment is important to block occupied core states to the 
valence neutrons. That is, Pauli
blocking is needed to remove forbidden states, which would
disappear under antisymmetrization. This can be taken into account by
several methods. In this work, a ``repulsive core'' in the $s$-wave component  
of the $\alpha$-$n$ subsystem is introduced with the requirement that the 
experimental phase shifts are correctly calculated. This method is referred 
in the literature as the PC method~\cite{IJThompson00}. 

The radiative capture of two neutrons by an alpha particle producing
$^6$He is dominated by a dipolar process from the $1^-$ continuum of
$^6$He to the $0^+$ ground state~\cite{RdDiego10}. 
For $^6$He, low-energy dipolar resonances have not been observed, then the electric dipole dominates over the magnetic dipole. A low-energy quadrupole resonance  does exist (as mentioned above). We have calculated both dipolar and quadrupolar electric contributions, concluding the quadrupole is several orders of magnitude lower than the dipole.
This means that the
reaction rate for this capture process is mainly governed by the dipolar
electric transition distribution $dB(E1)/d\varepsilon$ of
$^6$He. Then, to compute this distribution we need to
generate the THO basis for states $0^+$ and $1^-$. The $0^+$ THO basis
must provide a well-converged ground state. The $1^-$ THO basis must
have enough states close to the break-up threshold to get a smooth and 
detailed $B(E1)$ distribution in that region. 
Using the parameters $b$ and $\gamma$ from the analytical LST one can find
the most suitable THO basis for each total angular momentum ($0^+$ and
$1^-$).  The THO bases were truncated at maximum hypermomentum $K_{max}=20$, 
as it was sufficient to have a good description of the system and provide 
converged results.

  \subsection{States $j^{\pi}=0^+$}

The $0^+$ states are described with an analytical THO basis defined  by
parameters $b=0.7$ fm and $\gamma=1.4$ fm$^{1/2}$, trying to minimize the size
of the basis needed to reach convergence of the ground state. We found that a
basis with larger $\gamma/b$ has a too large energy distribution to provide a
fast convergence for the ground state. On the other hand, a basis with smaller
$\gamma/b$ has a very large hyperradial extension  and does not describe
properly the interior region of the potential where the ground state probability
is larger. The three-body force parameters are taken as $v_{3b}=-2.45$ MeV, $r_{3b}=5$ fm, and $a_{3b}=3$ in order to adjust both, the ground-state energy and the matter radius of $^6$He.

\begin{figure}
\includegraphics[width=0.95\linewidth]{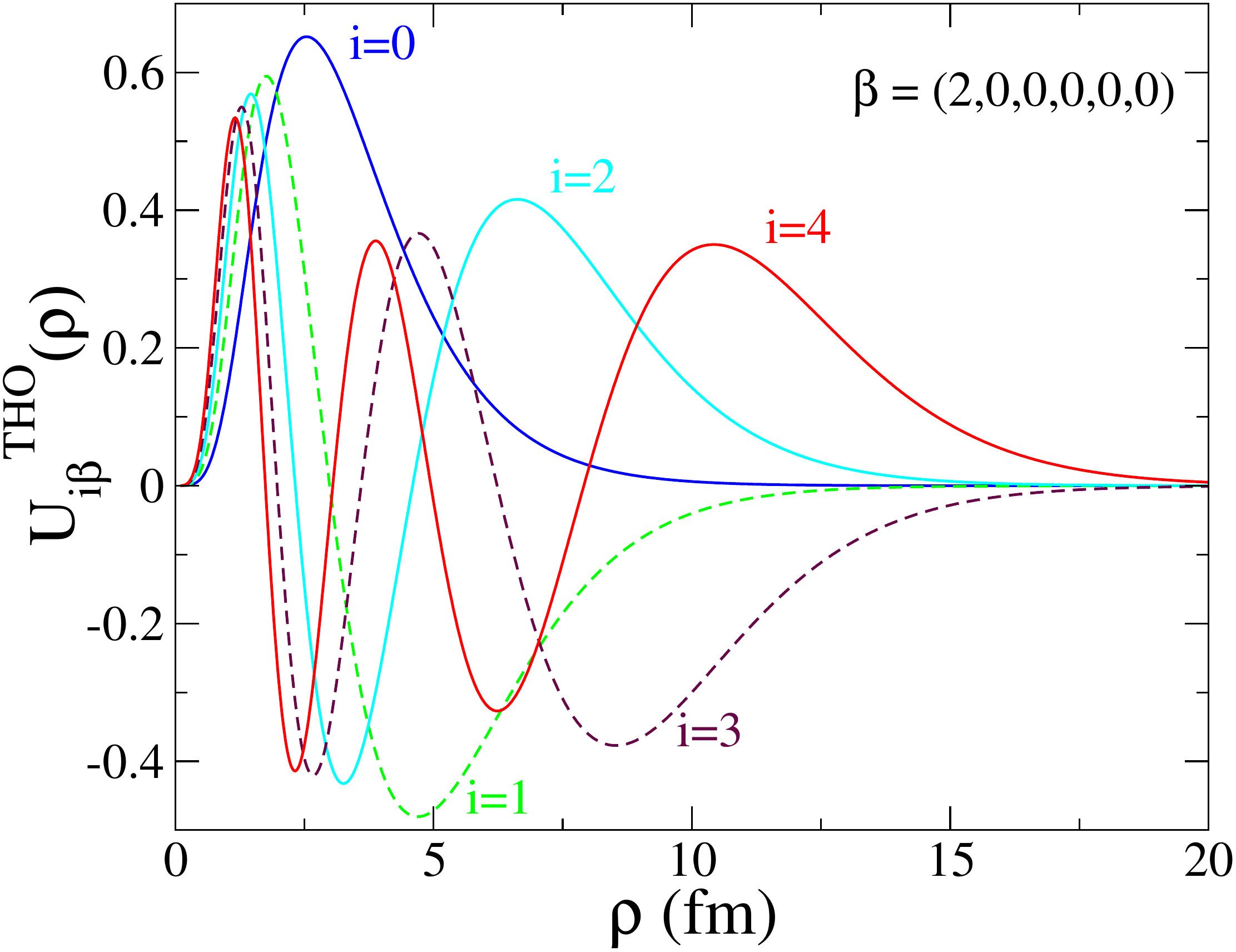} 
\caption{(Color online) First five THO hyperradial wave functions for the channel 
$\beta\equiv\{2,0,0,0,0,0\}$, the most important channel in the g.s. wave function.}
\label{fig:THO}
\end{figure}

In Fig. \ref{fig:THO} we show the first THO hyperradial wave functions for the 
channel $\beta\equiv\{2,0,0,0,0,0\}$, using the given LST and three-body force parameters. This channel is
the most important ground-state channel, with a 78.6\% contribution to the total
norm. We can see in the figure that as $i$ increases, the functions are more
oscillatory and explore larger distances. 

\begin{figure}
\includegraphics[width=0.95\linewidth]{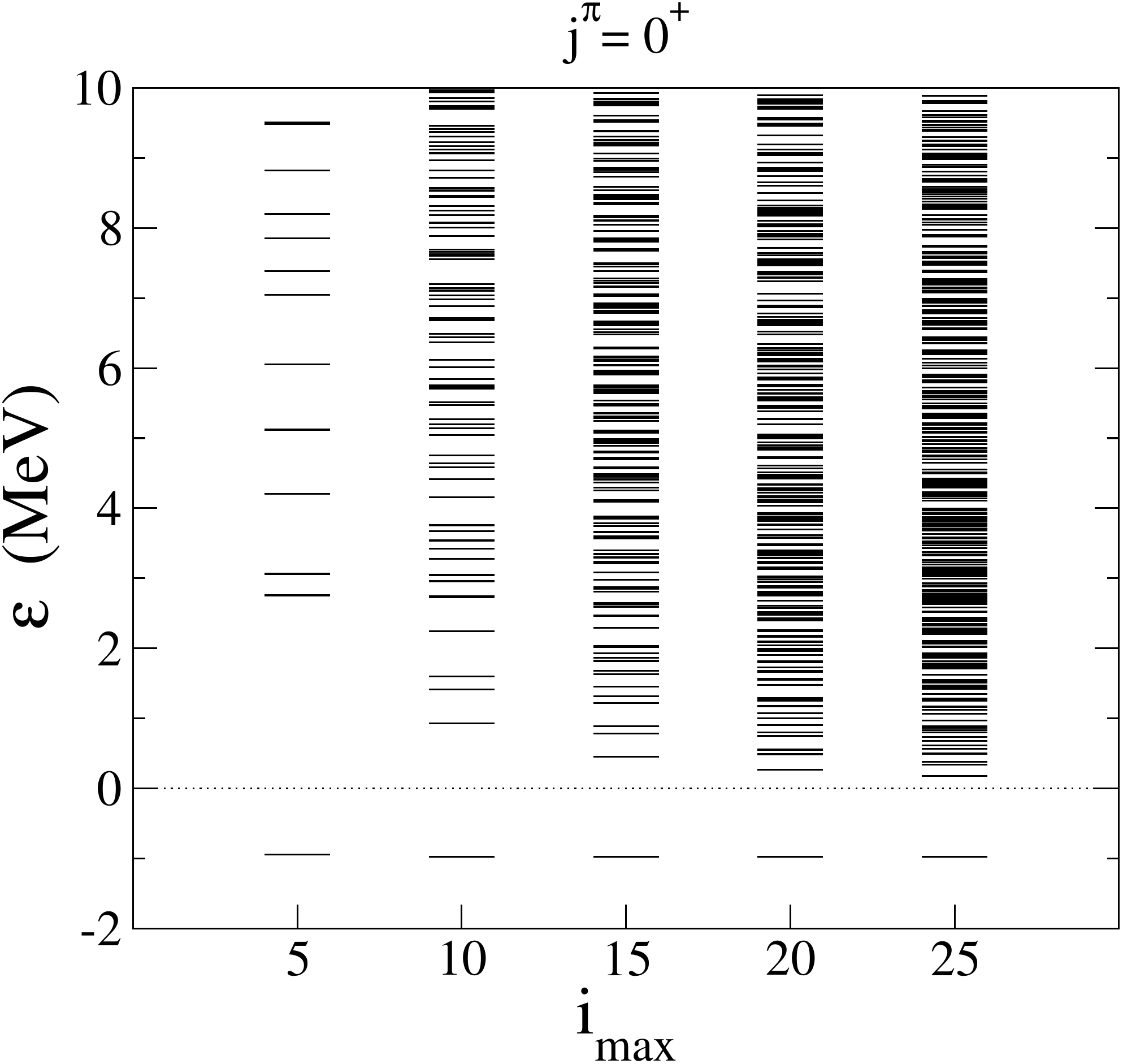}
\caption{Eigenvalues for $j^\pi=0^+$ up to 10 MeV.}
\label{fig:spectra0}
\end{figure}

In Fig. \ref{fig:spectra0} the Hamiltonian eigenvalues for $j^\pi=0^+$,  for
an increasing number of hyperradial excitations, $i_{\rm max}$, are presented up to
10 MeV. The calculated ground-state is stable, has a binding energy of
0.9749 MeV and a rms point nucleon matter radius of 2.554 fm. Calculations
assume an $\alpha$ radius of 1.47 fm.
In Table I the ground-state energy $\varepsilon_B$ and matter radius $r_{mat}$ 
are shown as a function of the maximum number of hyperradial excitations
$i_{\rm max}$. We observe a fast convergence of this two  ground-state observables
within this THO basis.

\begin{table}
 \begin{tabular}{ccc} 
 \toprule
 $i_{\rm max}$ & $\varepsilon_B$ (MeV) & $r_{mat}$ (fm)\\
 \colrule
 5  & $-$0.9452 & 2.511\\
 10 & $-$0.9744 & 2.552\\
 15 & $-$0.9748 & 2.554\\
 20 & $-$0.9749 & 2.554\\
 25 & $-$0.9749 & 2.554\\
 \botrule
 \end{tabular}
\caption{Ground-state energy $\varepsilon_B$ and matter radius $r_{mat}$
 as a function of $i_{\rm max}$. A fast convergence is observed.}
\end{table}

\begin{figure}
\vspace{0.1cm}
\includegraphics[width=\linewidth]{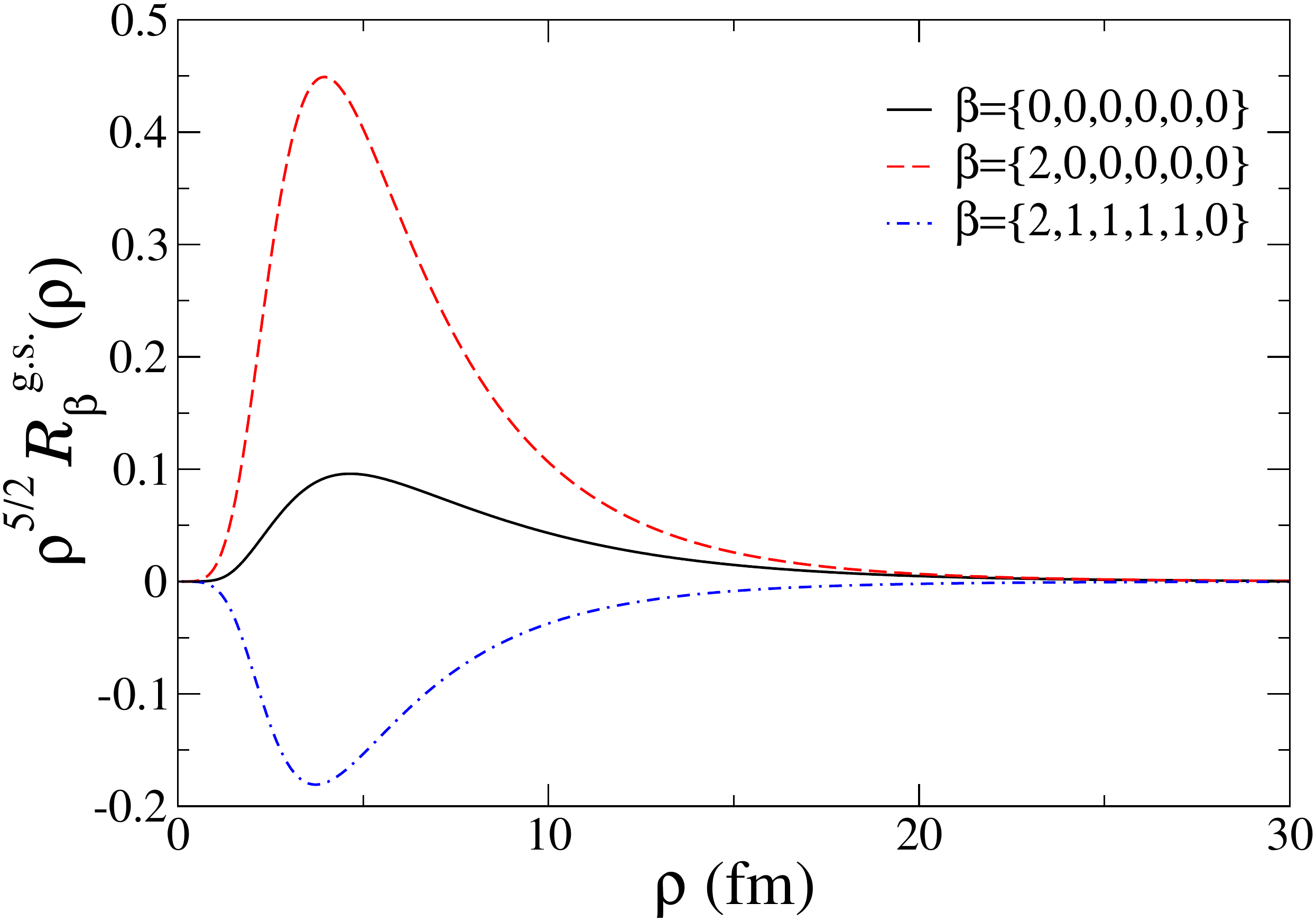}
\caption{(Color online) Hyper-radial wave function,
$\mathcal{R}_\beta^{\text{g.s.}}(\rho)$,  for the first three channels included
in the ground state of $^{6}$He.}
\label{fig:wf}
\end{figure}

The first three hyperradial components of the ground-state 
wave function for $i_{\rm max}=25$ are presented in Fig. \ref{fig:wf}. The
curves  match a reference calculation of the ground-state wave function corresponding to the same model Hamiltonian. By reference calculation we mean the procedure presented in Ref.~  \cite{IJThompson00} and implemented in the codes FaCE~\cite{IJThompson04} and sturmxx~\cite{sturmxx}, using a suitable basis for bound states, the so-called Sturmian basis.

Once the $0^+$ ground state is obtained, the $1^-$ states in the continuum have
to be generated. However, no reference is available to fix the $1^-$ three-body
force. For the $2^+$ continuum states there is a resonance
experimentally observed at 0.824 MeV over the breakup threshold. Thus,
usually the three-body force is fixed to set the $2^+$ resonance at the
experimental value and it is accepted the same three-body force for the
$1^-$ states. So we generate first the THO basis for 2$^+$
states and adjust the position of the $2^+$ resonance by using a particular
three-body force. Then we use the same force for the $1^-$ states.

  \subsection{$\boldsymbol{2^+}$ states}

%
The $2^+$ states were described with a basis defined by $b=0.7$ fm and 
$\gamma=2.0$ fm$^{1/2}$. This basis has a small hyperradial extension and
spreads the eigenvalues obtained upon diagonalization at higher energies. This
choice allows us to have only one pseudo-state presenting the characteristics of
the resonance, since the rest of states are sufficiently above the resonance
energy position for medium-size bases. In this way we can adjust the resonance
energy, setting the energy of this state to the experimental value.
Then, the three-body force parameters are taken as $v_{3b}=-0.90$ MeV, $r_{3b}=5$ fm, and $a_{3b}=3$. 

 \begin{figure}
\includegraphics[width=0.95\linewidth]{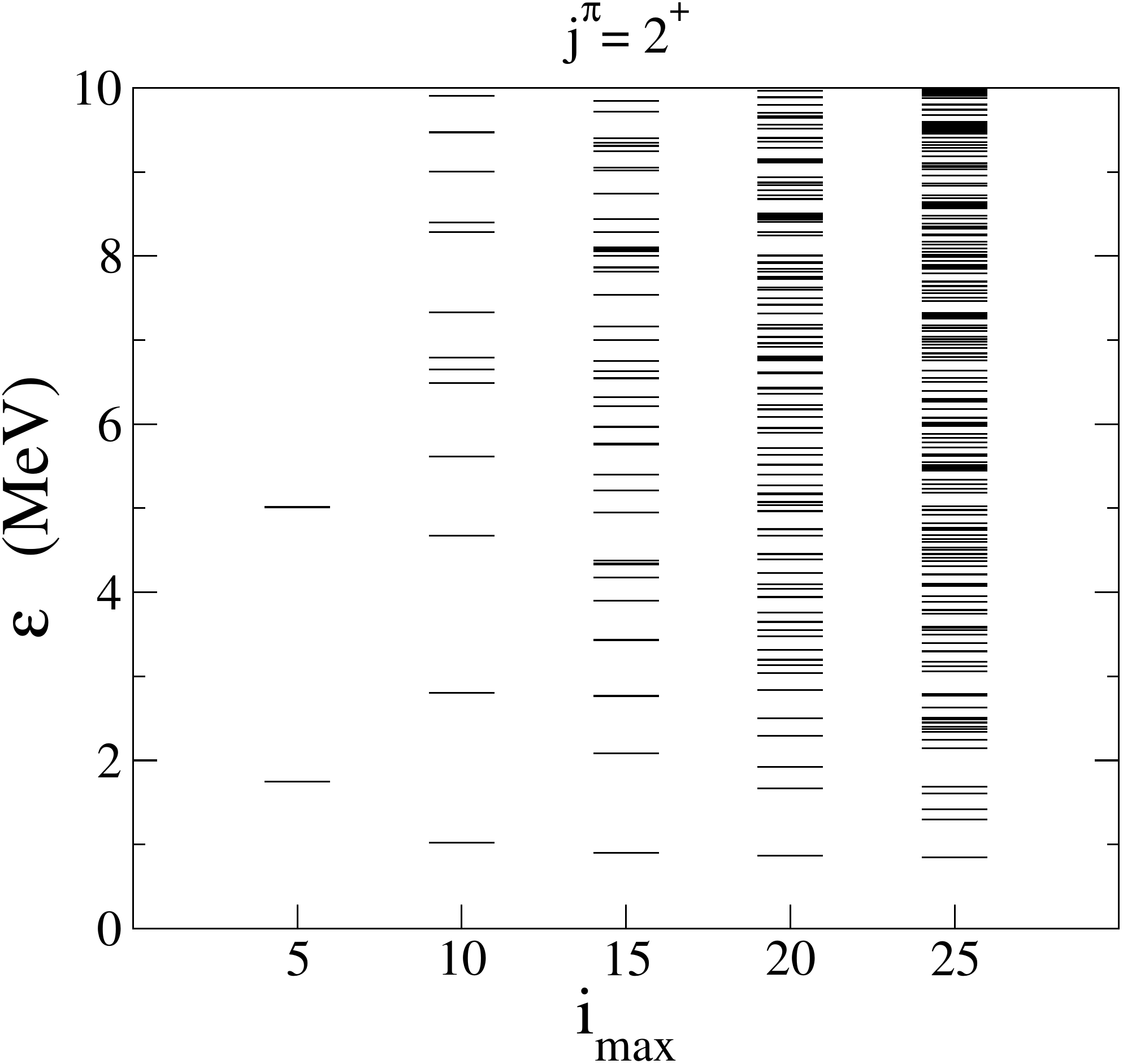}
\caption{Eigenvalues for $j^\pi=2^+$ up to 10 MeV.}
\label{fig:spectra2}
\end{figure}

\begin{figure}
\includegraphics[width=0.95\linewidth]{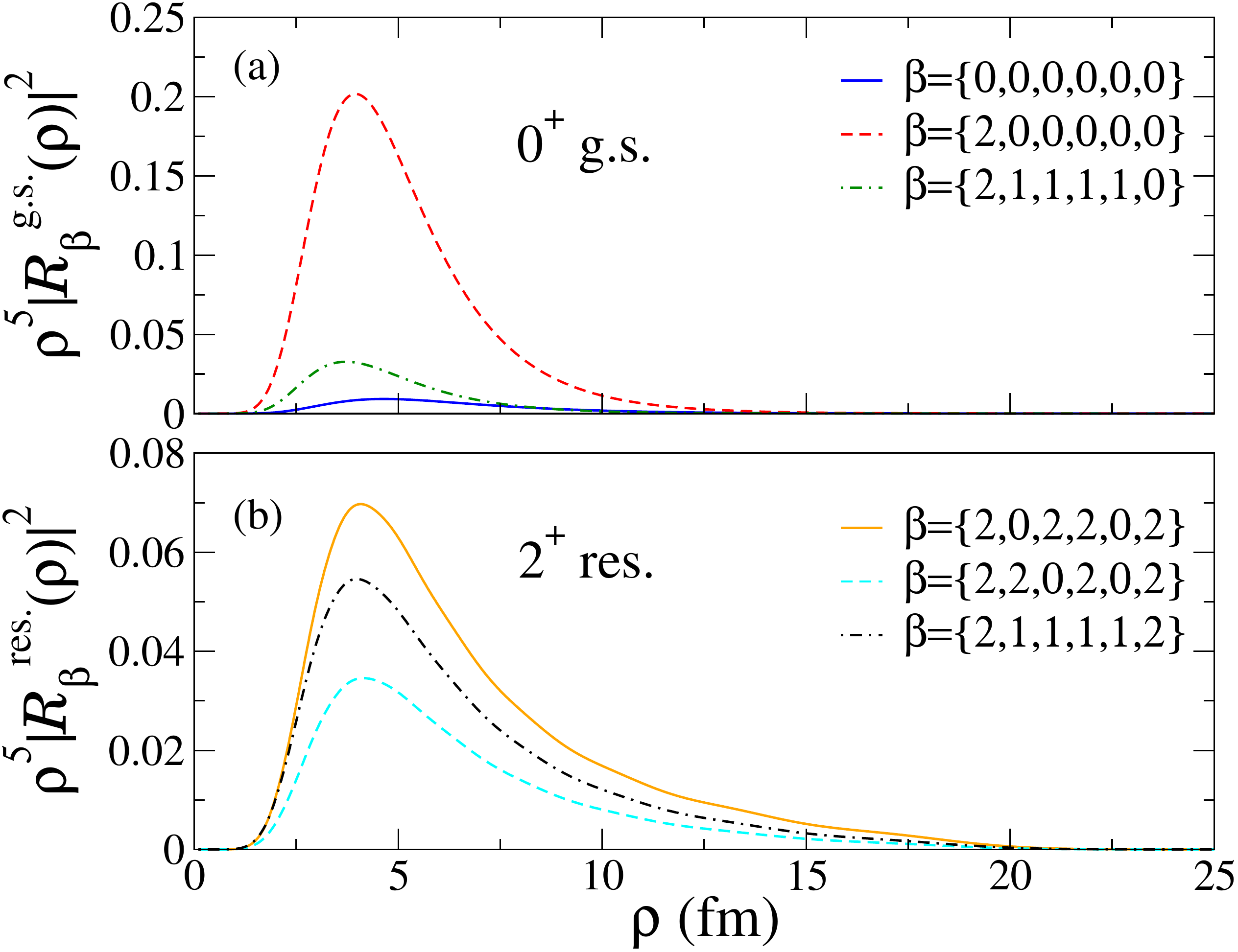}
\caption{(Color online) Ground-state (a) and resonance state (b)  probabilities.}
\label{fig:reswf}
\end{figure}

In Fig. \ref{fig:spectra2}, the eigenvalues of the Hamiltonian for $j^\pi=2^+$ 
states, for an increasing number of hyperradial 
excitations, are shown. The lowest state is rather stable and close to the
energy  of the known $2^+$ resonance, 0.824 MeV. 
In Fig.~\ref{fig:reswf}, we present the probability density for this 
first $2^+$state, compared with the $0^+$ ground state probability. 
The contributions of the three most important channels for each one are shown. 
We can see the PS
representing the resonance is a state with a large probability in the interior
part, similar to a bound state. 

\subsection{$\boldsymbol{1^-}$ states}

The preceding calculation on $2^+$ states with the low-lying resonance as
reference allows us to select the three-body force  ($v_{3b}=-0.90$ MeV, $r_{3b}=5$ fm, $a_{3b}=3$) to be included in the
calculation of the required $1^-$ states.

\begin{figure}
\includegraphics[width=0.95\linewidth]{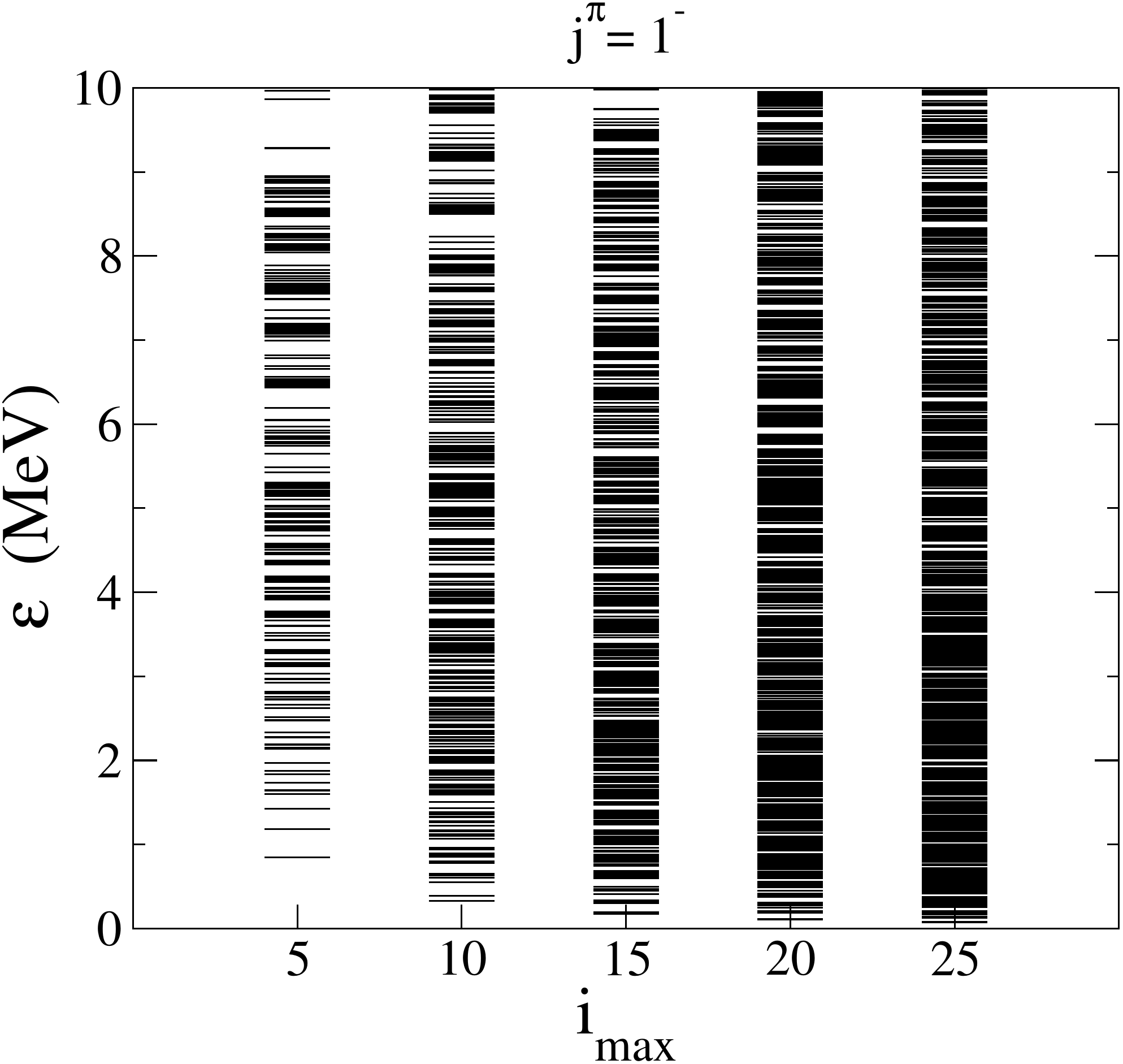}
\caption{Eigenvalues for $j^\pi=1^-$ up to 10 MeV.}
\label{fig:spectra1}
\end{figure}

To get a well-defined $B(E1)$ distribution near the origin, we need, for $1^-$ 
states, a basis  which has a large hyperradial extension to concentrate many
eigenvalues close to the breakup threshold.
For this purpose we use a THO basis with $b=0.7$ fm and $\gamma=1.0$ fm$^{1/2}$.
The eigenvalues of the Hamiltonian for $j^\pi=1^-$ states are presented for
different $i_{\rm max}$ values in Fig. \ref{fig:spectra1}. If we compare this $1^-$
spectrum with the $0^+$ and $2^+$ spectra for a fixed $i_{\rm max}$, it is clear the
difference in eigenstates density depending on the extension of the basis, that
is, depending on the LST parameters $b$ and $\gamma$.

Next we can calculate the discrete  transition probabilities $B(E1)$ 
from the $0^+$ ground state to the $1^-$ eigenstates. We have first checked the completeness of the basis for a
given $i_{\rm max}$ comparing the sum of the discrete $B(E1)$ transition
probabilities with the sum rule Eq.~(\ref{eq:sumrule}). This is given in Table
II. The summation 
converges to the exact value given by the sum rule, 1.493 e$^2$fm$^2$.
\begin{table}[!hb]
 \begin{tabular}{ccc} 
 \toprule
 $i_{\rm max}$ & & $\sum B(E1) $ (e$^2$fm$^2$) \\
 \colrule
  5 & & 1.402  \\
 10 & & 1.489  \\
 15 & & 1.492  \\
 20 & & 1.492  \\
 25 & & 1.493  \\
 30 & & 1.493  \\
 35 & & 1.493  \\
 \botrule
 \end{tabular}
\caption{Sum of $B(E1)$ as a function of $i_{\rm max}$.}
\end{table}

\begin{figure}
\includegraphics[width=0.95\linewidth]{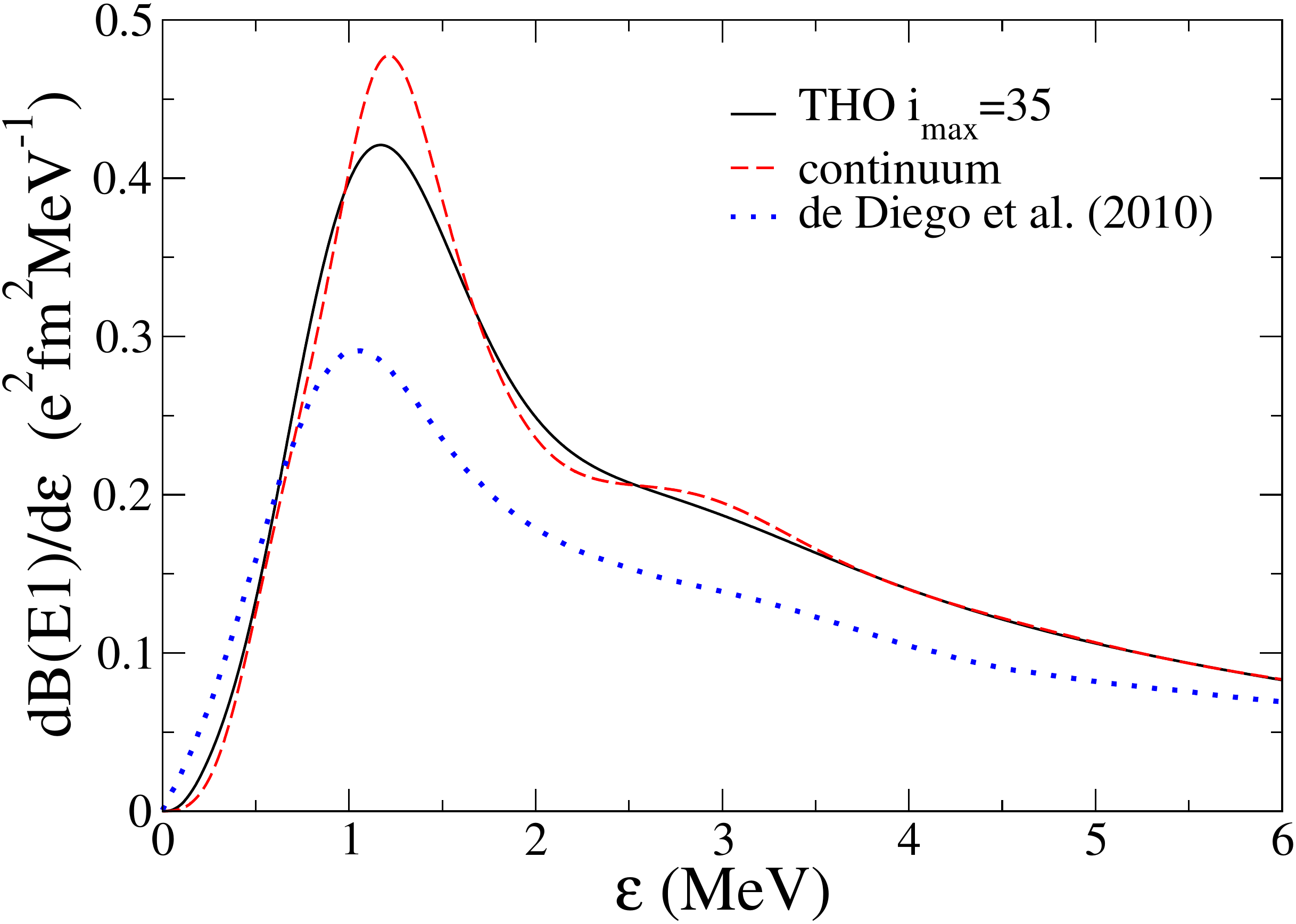}
\caption{(Color online) $B(E1)$ distribution up to 6 MeV: this work (full
black line), a calculation using the actual continuum wave functions, that in
this case can be calculated, (broken red line), and Ref. \cite{RdDiegoTh}
(dotted blue line).}
\label{fig:dbe1}
\end{figure}

For the evaluation of the transition probabilities, we use a THO basis with $i_{\rm max}=35$ in order to  obtain
a detailed behavior for the low energy part of the $B(E1)$ distribution.
In Fig.~\ref{fig:dbe1} we show, up to 6 MeV, a reference calculation
obtained by using the actual three-body continuum wave functions which, in
this simple case, can be computed easily~\cite{IJThompson00} (dash red line).  
To generate the continuum wave functions we have used the codes FaCE~\cite{IJThompson04} and sturmxx~\cite{sturmxx} with the same model Hamiltonian.
If the smoothing of our THO calculation is done using the overlap with the continuum wave functions, the obtained $B(E1)$ distribution is indistinguishable from the reference one. This guarantees that the formalism presented here is working correctly.
However, since our interest is to extend this formalism to other systems for which the true continuum wave functions are difficult to obtain, we propose an alternative smoothing procedure following Eqs.~(\ref{eq:poisson}) and (\ref{eq:distBE}). In Fig.~\ref{fig:dbe1} the THO distribution for $B(E1)$, using this alternative smoothing, is shown  (full black line). We have used Poisson distributions  with parameter $w=30\sqrt{\varepsilon_n}$,  such that it  ensures a smooth $B(E1)$ distribution without spreading it unphysically. Due to the large number of basis states we have near  the threshold, the energy dependence of $w$ is convenient to produce a smooth distribution in that region.
 The total $B(E1)$ strength is the same for both
calculations (solid and dashed lines) and the behavior is similar, although small differences are observed in the medium energy range.

It is also included in Fig.~\ref{fig:dbe1} a calculation taken
from Ref. \cite{RdDiegoTh}. In that work, the hyperspherical adiabatic expansion method is used instead of the HH method. Then, the three-body states are calculated by box boundary conditions, obtaining a discrete spectrum. The discrete $B(E1)$ values are smoothed using the finite energy interval approximation.    
This calculation clearly have a different behavior
at low energies. The difference comes from the difficulty to have a large energy level density at low energies solving the problem in a box.
It is also apparent that the total $B(E1)$ from this
calculation is considerably lower than ours.
In our calculation the smoothed $B(E1)$ energy distribution is very
well-defined close to the break-up threshold since we have been able, using the
analytical THO, to build a basis for $1^-$ states concentrating many eigenvalues
close to the breakup threshold. In the literature, one can find other $B(E1)$
distributions for $^6$He using different three-body formalisms. We
would like to cite \cite{Myo01} and \cite{Baye09}, globally both compare
reasonably well with our results but have not been included in
Fig.~\ref{fig:dbe1} since it is not possible to extract from the
plots presented in those publications the detailed behavior at low
energies. Without this information, one cannot calculate converged
reaction rates below 1--2 GK. 

 It is worth mentioning that the available experimental data~\cite{Aumann99} (not shown in Fig.~\ref{fig:dbe1}) differ significantly from all published theoretical calculations. In particular the data do not show the enhancement at energies around 1 MeV. Either new experiment or reanalysis of the existing data is clearly needed.

\begin{figure}
\includegraphics[width=0.95\linewidth]{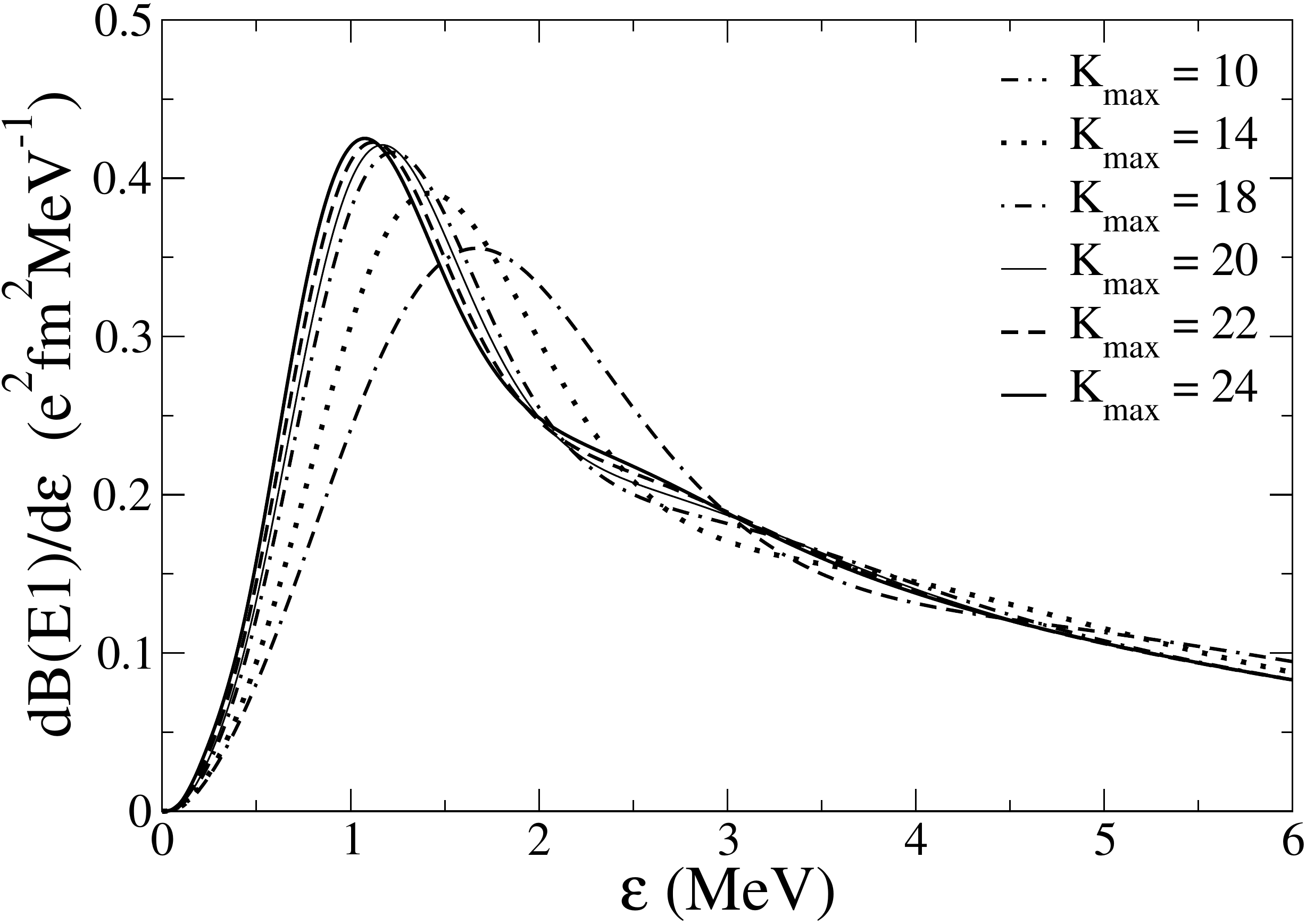}
\caption{$B(E1)$ distribution up to 6 MeV as $K_{\rm max}$ increases.}
\label{fig:dbeK}
\end{figure}

 In order to show the convergence of calculations with  $K_{\rm max}$ and that $K_{\rm max}=20$ is sufficient to provide converged results,  we present in Fig.~\ref{fig:dbeK} the $B(E1)$ distribution for different  $K_{\rm max}$ values. In these calculations the same two- and  three-body forces are kept fixed. 
It is clear from the figure that the calculations for  $K_{\rm max}=20$, 22, and 24 are very close together.

\begin{figure}
\includegraphics[width=0.95\linewidth]{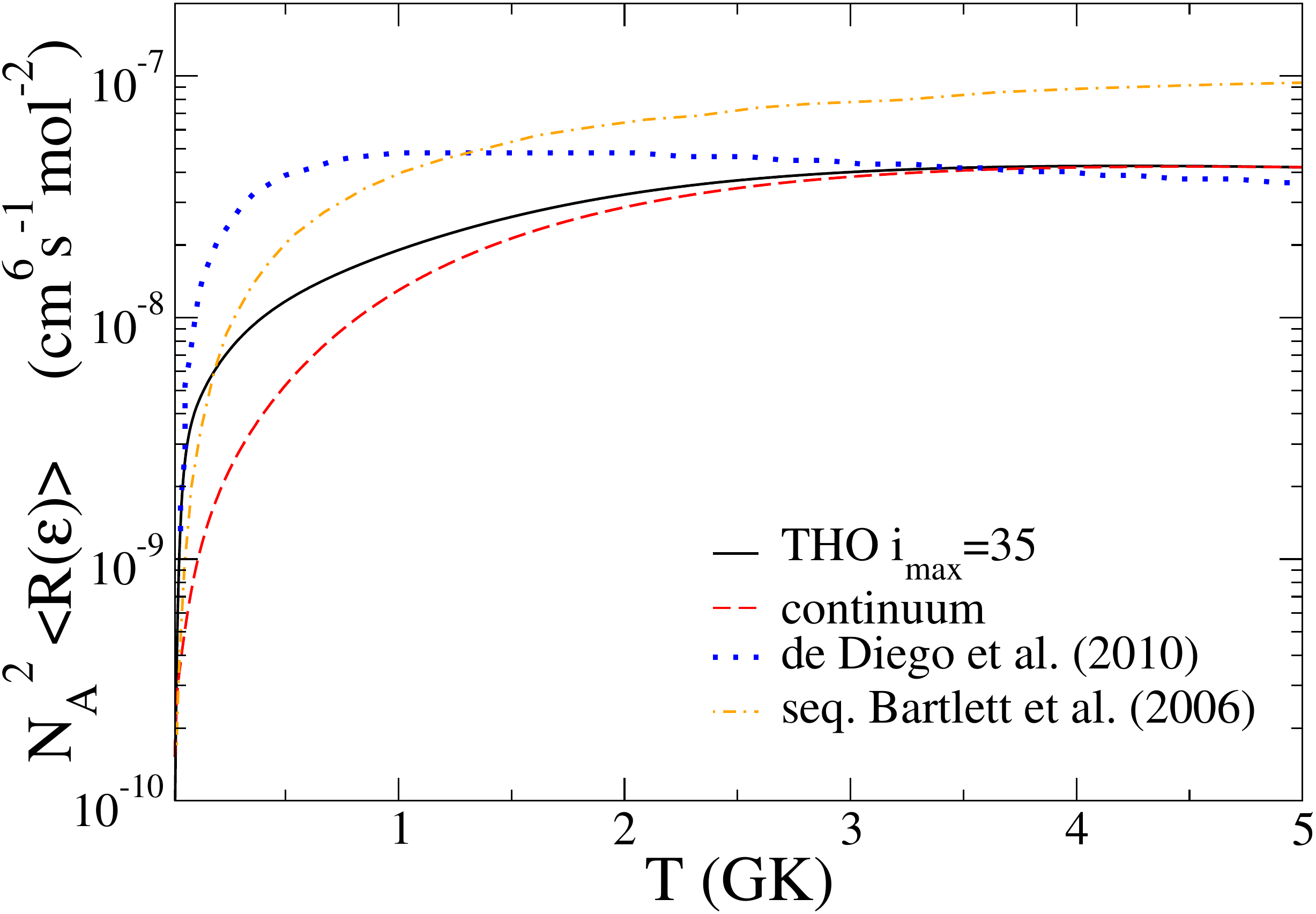}
\caption{(Color online) Reaction rate for the radiative capture
$\alpha+n+n\to^6\text{He}+\gamma$ with different models: this work (full black
line), a reference calculation using the actual three-body continuum
wave functions (dash red line), the results from Ref. \cite{RdDiegoTh}
(dotted blue line), and the results from a sequential calculation
~\cite{Bartlett06} (dot-dashed orange line).}
\label{fig:rate}
\end{figure}

Once obtained the $B(E1)$ energy distribution, we can finally calculate the 
reaction rate (Eq.~(\ref{eq:aRE})) for the radiative capture reaction
\mbox{$\alpha+n+n\to~^6\text{He}+\gamma$}. In Fig.~\ref{fig:rate}, we present the result for
the low temperature region of astrophysical interest (0-5 GK). Our calculation is
the full black line. In the
same figure the reaction rate obtained using the actual three-body
continuum wave functions and the corresponding $B(E1)$ is represented
with a dash red line. The  dotted blue line is the calculation of Ref.~\cite{RdDiegoTh}.  We can see from the figure that our calculation
agrees very well with the reference calculation for low and high temperatures.
In the region between 0.1 and 1.5 GK, there are differences at most by a 3
or 4 factor. These differences with respect to the reference (red dashed
line) calculation are more than one order of magnitude in the same
temperature region for the  calculation of Ref.~\cite{RdDiegoTh}. This is due to the already referred different behavior of the corresponding $B(E1)$ distributions at low energies (below 0.5 MeV). We have
checked that this region is crucial for the computation of the
reaction rates, especially at low temperatures (below 1--1.5 GK).
We have also checked that small differences in the $B(E1)$ distributions between 0.5 and 3.5 MeV do not affect the calculated reaction rate provided the same total strength.  

In Fig.~\ref{fig:rate}, we have also included the results from a sequential model for the radiative capture~\cite{Bartlett06} (dot-dashed orange). This calculation presents the same behavior as ours but is a factor of two larger above 0.2 GK. It is worth mentioning that this sequential calculation assumes first the formation of a dineutron, which is controversial, and then the capture of this by an $\alpha$ particle.
An alternative sequential
process, presented also in Ref.~\cite{Bartlett06}, starts from a
neutron capture by the $\alpha$ particle to 
give $^5$He followed by the capture of a second neutron. This provides
a reaction rate more than two orders of magnitude smaller in all
studied ranges of temperatures.

We would like to stress  our calculation is based on a full three-body model that makes no assumptions about the reaction mechanism. In this sense all the physical sequential processes are implicitly included.

\section{Summary and conclusions}\label{sec:conclusions}

We have extended the analytical THO method for the study of three-body systems.
There are several advantages of the analytical over the numerical THO method:
(1) The
previous knowledge of the ground-state of the system is not needed. (2) The
analytical transformation is easy to be implemented in programming languages. 
(3) The versatility of the LST depending on the parameters $b$ and $\gamma$, allows one to design the best basis for the observable under study.

We have applied the formalism to the well-known Borromean nucleus $^6$He.  This
nucleus can be described as an $\alpha$ particle and two valence neutrons. We have
seen that the use of the analytical THO method allows a specific basis 
selection depending on the needs for each
angular momentum of the system and on the observable under study. We have
calculated a well-converged  $0^+$ ground state  and a rather stable $2^+$  resonant state. For $1^-$ states we have chosen a basis concentrating
many energy levels close to the breakup threshold in order to have a fine 
description for that region.

With these ingredients we have computed the $B(E1)$ transition probabilities 
from the $0^+$ ground state to the $1^-$ states. We have checked that the smoothing, using the overlap with the actual  continuum wave functions, produce the same $B(E1)$. The smoothing using Poisson distributions produces a similar result with small differences in the medium-energy region. In this case, the obtained  $B(E1)$ distribution
is well-defined at low energies (below 0.5 MeV), which is crucial to estimate properly observables
such as the reaction rate of the radiative capture \mbox{$\alpha+n+n\to~^6\text{He}+\gamma$}.

We have calculated the reaction rate of the radiative capture 
\mbox{$\alpha+n+n\to~^6\text{He}+\gamma$} from the $B(E1)$ distribution for temperatures of astrophysical interest. The result with Poisson smoothing for the $B(E1)$ provides a reasonable approach to the  continuum reaction rate. However it differs by a factor of 2 from  the  sequential mechanism presented in Ref.~\cite{Bartlett06}, which assumes the dineutron preformation, what is controversial. 
The differences with the reaction rate calculated in Ref.~\cite{RdDiegoTh}, using also a full three-body model, come from the different behaviors at low energies of the $B(E1)$ distributions (below 0.5 MeV) and the different total $B(E1)$ strengths.

The present results encourage the application of this formalism to more interesting astrophysical  
cases, such as $^9$Be, the triple-$\alpha$ process to produce $^{12}$C, or $^{17}$Ne. In the 
study of these systems,
one of the major problems is the proper treatment of the Coulomb 
interaction at large distances. However, this problem is absent in the PS 
methods, such as the analytical THO presented here. 

\begin{acknowledgments}
 Authors are grateful to  P. Descouvemont, R. de Diego,  E. Garrido, and I. J. Thompson for useful discussions and suggestions. 
This work has been partially supported by the Spanish Ministerio de
Econom\'{\i}a y  
Competitividad under Projects FPA2009-07653 and FIS2011-28738-c02-01, by Junta
de Andaluc\'{\i}a under group number FQM-160 and Project P11-FQM-7632, 
and by the Consolider-Ingenio 2010 Programme CPAN (CSD2007-00042). J. Casal 
acknowledges a FPU research grant from the Ministerio de Educaci\'on, Cultura y
Deporte.
\end{acknowledgments}

\bibliography{./bibfile}

\end{document}